\input{aipcheck}

\documentclass[
    ,final            
  ]
  {aipproc}

\layoutstyle{6x9}

\usepackage{amsfonts}
\def\ka{\kappa}
\newcommand{\su}{\mathfrak{su}}

\newcommand{\so}{\mathfrak{so}}
\newcommand{\SU}{\mathrm{SU}}
\newcommand{\AN}{\mathrm{AN}}

\newcommand{\ISU}{\mathrm{ISU}}
\newcommand{\SO}{\mathrm{SO}}

\newcommand{\Spin}{\mathrm{Spin}}

\def\kk{{\cal K}}
\newcommand{\R}{\mathbb{R}}

\newcommand{\N}{\mathbb{N}}

\newcommand{\be}{\begin{equation}}
\newcommand{\ee}{\end{equation}}
\newcommand{\bes}{\begin{eqnarray}}
\newcommand{\ees}{\end{eqnarray}}

\begin{document}

\title{The group field theory approach to quantum gravity: some recent results}


\author{Daniele Oriti}{
  address={Max Planck Institute for Gravitational Physics (Albert Einstein Institute)
\\ Am M\"{u}hlenberg 1
D-14476 Golm,
Germany, EU}
}

\begin{abstract}
We introduce the key ideas behind the group field theory approach to quantum gravity, and the basic elements of its formalism. We also briefly report on some recent results obtained in this approach, concerning both the mathematical definition of these models, and  possible avenues towards extracting interesting physics from them. \end{abstract}

\maketitle

\noindent The field of background independent quantum gravity is progressing fast \cite{libro}. Not only new research directions are being developed, and new important developments are taking place in existing approaches, but some of these approaches are converging to one another, leading to further progress. The group field theory formalism \cite{laurentgft,iogft,iogft2} can be understood in several different ways. It is a generalization matrix models for 2d quantum gravity \cite{mm}. It is an important part, nowadays, of the loop quantum gravity and spin foam approach to the quantization of 4d gravity \cite{carlo,SF}. It is a point of convergence of loop quantum gravity and of simplicial quantum gravity approaches, like quantum Regge calculus and dynamical triangulations \cite{iogft}. Recently, tools from non-commutative geometry have been introduced as well in the formalism, which, thanks to them, started to make tentative contact with quantum gravity phenomenology.

\noindent In this paper we introduce the general idea behind the GFT formalism, and some basic elements of the same (for more detailed introduction, we refer to \cite{iogft,iogft2, laurentgft}), then report briefly on some recent results. 

\section{The group field theory formalism}

\subsection{Motivation and key idea}

\noindent Group field theories are an attempt to define quantum gravity in terms of {\it combinatorially non-local quantum field theories on group manifolds}, related to the Lorentz or rotation group. Let us motivate briefly the three main elements in this characterization.

\

\noindent Quantum field theory is the best formalism we have for describing physics at both microscopic and mesoscopic  scales, from high energy particle physics to many-particle condensed matter physics. 

\noindent But can we still hope for a formulation of quantum gravity as a QFT, after the known failure of the attempt to formulate a consistent QFT of gravitons? And a more serious objection is that a good theory of quantum gravity should be background-independent, as it should explain origin and properties {\it of spacetime} itself, while we know how to formulate quantum field theories only on fixed backgrounds. Therefore, if an (almost) ordinary QFT it should be, quantum gravity can only be a QFT on some auxiliary, internal or \lq\lq higher-level\rq\rq space. 

\noindent We can then look at GR itself and try to identify some background (non-dynamical) structures that we could hope to provide such space. One is the internal, local symmetry group of the theory, i.e. the Lorentz group. This gives the primary motivation for using the Lorentz group (and related) in GFT. Another background structure is the configuration space of GR: the (meta-)space of (spatial) geometries on a given (spatial) topology, coined \lq\lq superspace\rq\rq by Wheeler. 
We will see that these two possibilities for the field domain in GFT actually coincide, because, as we have learned from loop quantum gravity, the set of possible geometries can be characterized in terms of (Lorentz) group elements. 

\noindent When suggesting a QFT for the microstructure of space, another natural question is: a QFT of which fundamental quanta? Again, we know that these cannot be gravitons. They have to be quanta {\it of space} itself, excitations around a vacuum that corresponds to the {\it absence of space}. GFT incorporates, in the choice of these quanta, ideas from both loop quantum gravity and simplicial quantum gravity, as we are going to see. 

\noindent The last ingredient of our {\it combinatorially non-local quantum field theory on group manifolds} that we have to introduce and motivate is the {\it non locality}. 

\noindent Consider a point particle in 0+1 dimensions, with action $S(X) = \frac{1}{2} X^2\,+\,\frac{\lambda}{3} X^3$. This action defines a trivial dynamics (for a trivial system), of course. What interests us here, however, is the combinatorial structure of its \lq\lq Feynman diagrams\rq\rq, i.e. the graphs that can be used as a convenient book-keeping tool in computing the corresponding partition function $Z = \int dX\, e^{- S_{\lambda}(X)}$ perturbatively in $\lambda$. These are simple 3-valent (because of the order of the \lq\lq interaction\rq\rq) graphs.

\begin{figure}[here]
  \includegraphics[width=4cm]{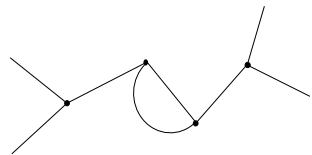}
  \caption{A Feynman graph for a point particle and the corresponding field theory}
\end{figure}

\noindent The fact that the Feynman diagrams of the theory are simple graphs like the above follows from 1) the {\it point-like} nature of the particle, and 2) the {\it locality} of the corresponding interaction, encoded in the identification of $X$ variables in the interaction term.

\noindent The same structure of diagrams is maintained, because the local nature of the interaction and the point-like nature of the corresponding quanta are maintained, also when moving to a field theory setting, i.e. going from the above particle dynamics to the corresponding field theory (still dynamically rather trivial), governed by the action: $S(\phi) = \frac{1}{2}\,\int dx\,\phi(x)^2\,+\,\frac{\lambda}{3}\,\int dx\,\phi(x)^3$. The associated Feynman amplitudes have integrations over position or momentum variables, but still, the combinatorics of the diagrams is the same.

\

\noindent Now we move up in combinatorial dimension. Instead of point particles, let us consider 1d objects, that could be represented graphically by a line, with two end points. We label these two end points with two indices $i,j$, and we represent the fundamental 1d objects by (e.g. orthogonal) {\it matrices} $M_{ij}$. We want to move up in dimension also in the corresponding Feynman diagrams. We want diagrams corresponding to 2-dimensional structures. For this, we have to drop the assumption of {\it locality}. We define an action for $M$, given by $ S(M)= \frac{1}{2}\,M_{ij}M_{ji}\, + \,\frac{\lambda}{3}\,M_{ij} M_{jk} M_{ki}$. The racing of indices $i,j,k$ in the kinetic and vertex term represent identification of the points labeled by the same indices. This graphical representation of the Feynman diagrams used in evaluating the partition function $Z = \int \mathcal{D}M_{ij} \, e^{- S(M)}$ gives 2-dimensional simplicial complexes of arbitrary topology, because obtained by arbitrary gluing of lines to form triangles (in the interaction vertex) and of these triangles to one another along common lines (as dictated by the propagator). 

\begin{figure}[here]
\includegraphics[width=6cm, height=2cm]{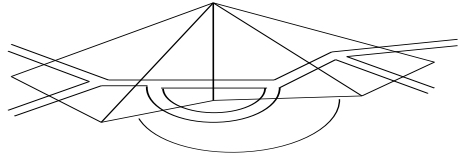}
\caption{A (piece of) Feynman diagram for a matrix model, of which we give both direct and dual (simplicial) representation; the two parallel lines of propagation correspond to the two indices of the matrix; the extra line on the bottom indicates identification of the two edges of the triangles.}
\end{figure}    

\noindent Matrix models have been quite successful in describing 2d quantum gravity \cite{mm}.

\

\noindent There is no obstruction to keep moving upward in combinatorial dimension. We can move from 1d objects represented by matrices, with indices labeling the end points of the line, to 2d (closed) objects, represented by tensors, with indices labeling the boundary edges of the same 2d objects. The simplest combinatorial choice is that of triangles (2d simplices) and thus of tensors $T_{ijk}$, with indices (representing the edges of the triangles) traced out in the interaction term in such a way as to represent 3d objects, tetrahedra (3-simplices) bounded by such triangles. 

\begin{figure}[here]
\includegraphics[width=4cm]{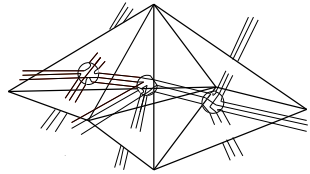}
\caption{A (piece of) Feynman diagram for a tensor model, of which we give both direct and dual (simplicial) representation; the three parallel lines of propagation (dual to the three edges in the triangles) correspond to the three indices of the tensor.}
\end{figure}   

\

\noindent The process of combinatorial generalization can be continued to tensor models whose Feynman diagrams are d-dimensional simplicial complexes. Moreover, while maintaining the combinatorial structure of the theory, we can generalize matrix and tensor models in the direction of adding degrees of freedom, i.e. defining corresponding field theories. The tensor indices will be replaced by variables taking values in appropriate domain spaces. Choosing these spaces to be group manifolds, we obtain group field theories.

\

\noindent A prototype GFT generalizing a tensor model for $T_{ijk}$ is, for example, the so-called Boulatov model: $$S(\phi)\,=\,\frac{1}{2}\int [dg] \phi(g_1,g_2,g_3)\phi(g_3,g_2,g_1)\,+\,\frac{\lambda}{4!}\,\int [dg] \,\phi(g_1,g_2,g_3)\phi(g_3,g_4,g_5) \phi(g_5,g_6,g_1)\phi(g_6,g_4,g_2) . $$

\subsection{The general formalism}
\noindent The GFT field
is a $\mathbb{C}$-valued function of D group elements
$\phi(g_1,..,g_d)$, for a group $G$, usually the
$SO(d-1,1)$ Lorentz group (or $SO(d)$ for Riemannian gravity). It is interpreted as the fundamental building block of quantum space, a second quantized
(d-1)-simplex, and each argument corresponds to one of its
boundary (d-2)-faces. One imposes invariance under diagonal action
of the group $G$: $\phi(g_1,...,g_d)=\phi(g_1g,...,g_dg)$ to give closure of the d (d-2)-faces to form a
(d-1)-simplex. The
mode expansion gives:
$$\phi(g_i)=\sum_{J_i,\Lambda,k_i}\phi^{J_i\Lambda}_{k_i}\prod_iD^{J_i}_{k_il_i}(g_i)C^{J_1..J_d\Lambda}_{l_1..l_d}, $$ with the $J$'s
labeling representations of $G$, the $k$'s being vector indices in the
representation spaces, and the $C$'s being group intertwiners, labeled by $\Lambda$. The GFT amplitudes will justify a geometric interpretation of the group variables as parallel transport of the gravity connection
along elementary paths dual to the (d-2)-faces, and of the representations $J$ as quantum numbers for volume operators for the same (d-2)-faces.

\begin{figure}[here]
  \includegraphics[width=10cm]{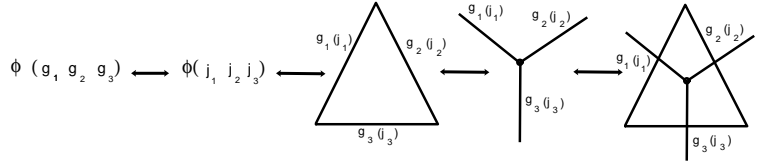}
\caption{The 3d example: the field, in group and representation picture, and its dual graphical representations as triangle and as spin network vertex}
\end{figure}

\noindent A simplicial space formed by
$N$ (d-1)-simplices is a product
of $N$ field operators, with contractions giving identifications of (d-2)-faces. GFT states are then, in representation variables, {\it spin
networks} of $G$. The combinatorics of the field arguments in the action 
$$
S= \frac{1}{2}\int
  dg_id\tilde{g}_i\,
  \phi(g_i)\mathcal{K}(g_i\tilde{g}_i^{-1})\phi(\tilde{g}_i)
  +
  \frac{\lambda}{(d+1)!}\int dg_{ij}\,
  \phi(g_{1j})...\phi(g_{d+1 j})\,\mathcal{V}(g_{ij}g_{ji}^{-1}),
$$
is constructed so to obtain d-dimensional complexes as Feynman diagrams. It describes the interaction of d+1 (d-1)-simplices
to form a d-simplex by gluing them along their (d-2)-faces
(arguments of the fields). The interaction is
specified by the choice of function $\mathcal{V}$. The kinetic
term involves two fields each representing a given (d-1)-simplex
seen from one of the two d-simplices (interaction vertices)
sharing it. Thus the function $\mathcal{K}$
specifies how the geometric degrees of freedom are propagated across simplices.  

\begin{figure}[here]
  \includegraphics[width=8cm]{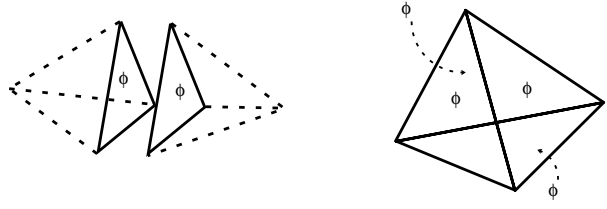}
  \caption{The 3d example: the simplicial representation of the interaction and of the propagation}
\end{figure}
\noindent Most of the work up to now has focused on the
perturbative aspects of quantum GFTs, using the expansion in
Feynman diagrams of the partition function:

$$ Z\,=\,\int
\mathcal{D}\phi\,e^{-S[\phi]}\,=\,\sum_{\Gamma}\,\frac{\lambda^N}{sym[\Gamma]}\,Z(\Gamma),
$$
where $N$ is the number of interaction vertices in the Feynman
graph $\Gamma$, $sym[\Gamma]$ is a symmetry factor for the graph
and $Z(\Gamma)$ the corresponding Feynman amplitude. 

\noindent Each edge of
the Feynman graph is made of $d$ strands, one for each argument of
the field, and each one is then re-routed at the interaction
vertex, following the pairing of field arguments in the vertex
operator. Each strand goes through several vertices, coming back
where it started, for closed Feynman graphs, and therefore
identifies a 2-cell. Each Feynman graph $\Gamma$ is then a cellular complex, that, because of the combinatorics of field
arguments in the action, is topologically dual to a
d-dimensional simplicial complex. 

\noindent The resulting complexes/triangulations can have arbitrary topology, because each simplicial complex is built up by arbitrary gluings of d-simplices to one another. 

\noindent In representation space, each Feynman diagram is
given by a spin foam (a 2-complex with faces $f$ labeled by
representation variables), and each Feynman amplitude by a spin foam model: $
Z(\Gamma)=\sum_{\{J_f\}} A(\{J_f\}).$ 

\noindent These models had been introduced in the loop quantum gravity approach to define the covariant dynamics of states of the quantum gravitational field, i.e. spin networks, and in fact transition amplitudes betwen given spin network states on the boundary is obtained in GFT
as the expectation value of field operators having the same
combinatorial structure of the two spin networks.
The same representation variables in loop quantum gravity,
have a geometric interpretation as quantum numbers of geometric operators, thus
each of these Feynman amplitudes defines a sum-over-histories for
discrete quantum gravity on the specific triangulation dual to the
Feynman diagram. They can be understood, therefore, as a new implementation of the idea of simplicial gravity path integrals, like quantum Regge calculus and dynamical triangulations. Actually, the amplitudes $Z(\Gamma)$ are usually at least directly motivated, from discretizations of classical gravity theories, and, in some models, take the explicit form of simplicial gravity path integrals. At the same time, this sum over geometries is generated within a sum over simplicial topologies corresponding to
the perturbative sum over Feynman diagrams. 

\

\noindent To summarize: GFT incorporates many ingredients of other approaches to
quantum gravity. Boundary states are spin networks, as in LQG, but they have a dual description as simplicial
spaces, as in simplicial quantum gravity. Their dynamics is
expressed as a covariant sum over quantum geometries as in spin foam
models, or a discrete gravity path integral, as in quantum Regge
calculus, but involves also a sum over inequivalent
triangulations, as in the dynamical triangulations approach.

\

\noindent As an explicit example, consider the Boulatov GFT model for 3d quantum gravity:
\bes S_{3d}[\phi]\,&=&\, \frac{1}{2}\int[dg]^3
\phi(g_1,g_2,g_3)\phi(g_3,g_2,g_1)\,+ \nonumber \\ &-&\,\frac{\lambda}{4!}\int [dg]^6
\phi(g_1,g_2,g_3)\phi(g_3,g_4,g_5)\phi(g_5,g_2,g_6)\phi(g_6,g_4,g_1) \label{boulatov}\ees
for a real field $\phi: \SU(2)^3\rightarrow \R$  invariant under the diagonal right action of
$\SU(2)$: $\phi(g_1,g_2,g_3)=\phi(g_1g,g_2g,g_3g)$ $\forall
g\in\SU(2)$. 

\noindent The Feynman diagrams are dual to 3d triangulations.
The Feynman amplitudes are $$
Z(\Gamma)\,=\, \prod_{L\in \Gamma} \int \,dh_L
\,\,\prod_{f}\,\delta (\prod_{L\in\partial f} h_L )\, 
$$
with a single delta function on the group $\delta (\prod_{L\in\partial f} h_{L} )$ for each 2-cell $f$ in the Feynman diagram, dual to a single edge $e$ of the simplicial complex, with argument given by the product of group elements $h_L$ each associated to a link in the boundary  of the 2-cell. 

\

\noindent The {\it same} amplitude can be given in representation space:
$$ Z(\Gamma)=\left(\prod_{f}\,\sum_{j_{f}}\right)\,\prod_{f}(2j_{f} +1)\,\prod_{v}\, \left\{ \begin{array}{ccc}
j_1 &j_2 &j_3
\\ j_4 &j_5 &j_6
\end{array}\right\} , $$
where the symbol for the vertex amplitude is the well-know $6j$-symbol from the recoupling theory of angular momentum
This is the Ponzano-Regge spin foam model.

\

\noindent At this stage the connection with simplicial path integrals is rather indirect, but still very clear. Consider the continuum 3d gravity action:
$$ S(e,\omega)\,=\,\int_{\mathcal{M}} tr \left( e\wedge F(\omega)\right) , $$  with variables the triad 1-form $e^I(x)\in su(2)$ and the 1-form connection  $\omega^{J}(x)\in su(2)$, with curvature $F(\omega)$. Introducing the simplicial complex $\Delta$ and its topological dual cellular complex, we discretize the triad in terms of Lie algebra elements associated to the edges of the simplicial complex as $E_e = E_{f} = \int_e e(x) = E^i J_i \in \su(2)$, and the connection in terms of parallel transports along links of the dual complex as $g_{L} = e^{\int_{L}\omega}\in \SU(2)$. The discrete curvature is given by: $G_{f} = G_e\,=\, \prod_{L\in \partial f} g_{L} = e^{F_{f}}\in \SU(2)$, and a discrete counterpart of the continuum action is: . $S(E,g)\, =\, \sum_{f\in\Gamma}\,tr\left(E_{f}\,G_{f}\right) $.

\noindent The path integral for this discrete theory on the simplicial complex $\Delta$ is {\it equal} to the previously computed GFT Feynman amplitude $Z(\Gamma$ for the diagram $\Gamma$ dual to $\Delta$:

$$Z(\Gamma)_{GFT} \, =\prod_{f}\int_{\su(2)} dE_{f}\prod_{L}\int_{\SU(2)}d g_{L} \,e^{i\,\sum_{f}\,tr\left(E_{f}\,G_{f}\right)}  \,=\,\prod_{L}\int_{\SU(2)}d g_{L} \,\prod_{f}\delta\left(G_{f}\right), $$
which, as we have seen, can be also expressed as a function of group representations $j_e$.

From all this, we also learn that spin foam models can be seen as a way of re-writing simplicial gravity path integrals expressed in connection (group) variables.

\section{Some recent results}

\subsection{New GFT (spin foam) models for 4d gravity}
\noindent The first task for a candidate approach to quantum gravity is the construction of interesting models for the microscopic quantum dynamics of space in 4 dimensions. 

\noindent The starting point of all recent model building has been the (Plebanski) formulation of 4d gravity as a constrained BF theory, in terms of $\so(4)$ Lie algebra-valued 1-form connection $\omega$ and  2-form $B$:
$$
S(\omega,B,\phi)=\int_{\mathcal{M}}\left[ B^{IJ}\wedge
F_{IJ}(\omega)-\frac{1}{2}\phi_{IJKL}B^{KL}\wedge B^{IJ}\right].
$$
Variations with respect to the Lagrange multiplier $\phi$, in fact, give constraints on the $B$ variables, whose solution \cite{SF} force $B$ to be a function of a tetrad field $e$: $B^{IJ}\,=\,\pm\,\frac{1}{2}\,\epsilon^{IJ} {}_{KL}\,e^{K}\,\wedge\,e^{L} $. On these solutions, the action becomes the Palatini action for gravity.

\

\noindent We know how to discretize and quantize BF theories in any dimension, and in particular we know how to construct the corresponding spin foam models and GFT action, and we also know how to discretize the Plebanski constraint on an arbitrary simplicial complex. The idea is then to start from such GFT action for BF: 

\bes S[\phi] &=& \frac{1}{2}\int dg_i \,[\phi(g_1,g_2,g_3, g_4)]^2 \, + \, \frac{\lambda}{5!}\int dg_j\,\left[\phi(g_1,g_2,g_3, g_4)
\phi(g_4,g_5,g_6, g_7)\right. \nonumber \\ &{}&\, \left.\phi(g_7,g_3,g_8, g_9)\phi(g_9,g_6,g_2,g_{10})\phi(g_{10},g_8,g_5,g_1)\right] , \ees

for a real field: $\phi(g_1,..,g_4): \SO(4)^{\times 4} \rightarrow \mathbb{R}$, symmetric under (local Lorentz): $\phi(g_1 g, g_2 g,..., g_4 g) = \phi(g_1,...,g_4)$,
or from the corresponding known quantum boundary states  ($\SO(4)$ spin networks) or Feynman (spin foam) amplitudes: 
\bes
Z(\Gamma) =\, \prod_{L} \int d g_{L}
\,\prod_{f}\,\delta (\prod_{L\in\partial f} g_{L} )\, = \sum_{\{j_+, j_- \}} \prod_f (2j_+ + 1) (2 j_- +1)\prod_v \, \big\{ 15j \big\}^v_+ \big\{ 15j \big\}^v_- , \ees 
(we used the selfdual/anti-selfdual splitting of $\SO(4)$ representations, and the vertex amplitude is the $15j$-symbol from the recoupling theory of angular momentum), which we know how to contruct from the same quantum states, and impose suitable restriction on its variables to impose the discrete Plebanski constraints.

\

\noindent Now we are faced with two possible strategies:
\begin{enumerate}
\item define a GFT model with Feynman amplitudes being manifestly simplicial path integrals for BF with gravity constraints; then re-write them as spin foam models;

\item find a quantum version of Plebanski constraints, impose them on quantum BF spin network states to get gravity spin network states, then construct spin foam/GFT amplitudes from these states; finally, check that they encode correctly simplicial geometry.

\end{enumerate}

\noindent The first strategy is made more difficult by the fact that we do not have a formulation of GFT reproducing a BF simplicial path integral with both $B$ and $g$ variables, and the gravity constraints should be imposed on $B$ variables. 

\ 

\noindent The second strategy is based on geometric quantization and stems from the identification of the $B$ variables, at the quantum level, with generators $T^{IJ}$ of the $\so(4)$ Lie algebra, which then act as operators on the $\SO(4)$ spin network states labelled by representation of the same group. The classical constraints, functions of the $B$'s are then also turned into operators and imposed at the quantum level on such spin networks, to give restrictions on both the representation attached to their links and on the intertwiners associated to their vertices \cite{EPR1,EPRL}. The resulting restricted spin networks are identified as gravity spin networks and used to construct new spin foam vertex amplitudes:
$$ A_v(k_f, i_e)=\sum_{i_a^-i_a^+}15j(j^+_{ab},i^+_{a})\ 15j(j^-_{ab},i^-_{a})\bigotimes_{a=1,5} f_{i_a^-i_a^+}^{i_a}, $$
where $j_{ab}^{\pm}=\frac{1\pm\gamma}2\  k$ are selfdual/anti-selfdual $\SU(2)$ representations, associated to the 2-cells of the Feynman diagram, $k_{ab}$ are representations of the diagonal $\SU(2)$ group sitting inside $\SO(4)$, $\gamma$ is an arbitrary constant (Immirzi parameter), $i_a^\pm$ are selfdual/anti-selfdual intertwiners associated to the links of the same diagram, $i_a$ are $\SU(2)$ intertwiners associated to the same links, and the functions $f$ are appropriate mapping coefficients from $\SO(4)$ to $\SU(2)$.
This is the EPR(L) spin foam model, currently the source of many new developments  \cite{EPR1,EPRL,FC}.
\footnote{A third strategy, closely related to the second, is based on the use of group coherent states and leads depending on how it is done, to the EPR(L) or to the so-called Freidel-Krasnov \cite{FK} model.}.
This model has two very nice properties: 1) its boundary spin networks match exactly those of canonical loop quantum gravity, with the same spectra for geometric operators; 2) the vertex amplitude reduces to the (cosine of) the Regge action in some semi-classical limit, confirming that the model correctly captures simplicial geometry at least for a single 4-simplex and at least in such approximation.

\

\noindent In order to follow the first strategy, it would be necessary to have a GFT depending on Lie algebra variables representing the $B$ variables of BF theory. It turns out that this requires a whole new representation of the GFT formalism \cite{ioaristide1}, using tools from non-commutative geometry \cite{laurentmajid}.

\

\noindent Consider the non-commutative $\su(2)$ Lie algebra, which is isomorphic to $\mathbb{R}^3$ as a vector space.
On such non-commutative space, one can consider functions on $\mathbb{R}^3$ endowed with a star product \lq\lq *\rq\rq  reflecting the non-commutativity of the underlying algebra. This is done first introducing
non-commutative plane waves depending on a group element $g\in\SU(2)$ and a Lie algebra element $x\in \su(2)$, as: $ e_g(x):  \su(2) \times \SU(2) \rightarrow \mathbb{C} : (x,g) \rightarrow \,e^{i \frac{1}{2}Tr(x g)}$.  
This leads to the definition of a non-commutative Fourier transform: $C(\SU(2)) \rightarrow  C_*(\mathbb{R}^3)\sim C(\su(2))$
$$
\phi(x)\, =\, \int_{\SU(2)} dg\, \phi(g) e_g(x)\,=\, \int_{\SU(2)} dg\, \phi(g)\, e^{i \frac{1}{2}Tr( x g)} ,
$$
with *-product defined on plane waves to respect the group multiplication, as: $\left(e_{g_1} * e_{g_2}\right)(x)\,=\, e^{i \frac{1}{2} Tr(x g_1)}\,e^{i \frac{1}{2} Tr(x g_2)}\, =\, e^{i \frac{1}{2} Tr(x g_1g_2)}\,=\, e_{g_1g_2}(x)$
and by linearity on generic functions.
The same NC Fourier transform can be inverted: $\phi(g)\,=\, \int d\vec{x}\, \left(\phi \, *\, e_{g^{-1}}\right)(x) $, modulo some technical subtleties.
The extension of this non-commutative Fourier transform to $\SU(2)^D$ and to arbitrary compact groups, including $\SO(4)$, is immediate.

\

\noindent The idea is then to use this transform to turn GFTs into {\it non-local and non-commutative field theories on Lie algebras}.

\

\noindent Let's first see what this gives in the 3d case. Using the Fourier transform:
\be
\phi(x_1, x_2,x_3) = \int dg_1\,..\,dg_3\, \phi(g_1,g_2,g_3)\, e_{g_1}(x_1)\,e_{g_2}(x_2)\,e_{g_3}(x_3) \,,\ee
we can re-write the whole action (\ref{boulatov}) for the Boulatov model in these new $x$ variables. 

\

\noindent For the Feynman amplitudes in this new representation, we find 
\be
Z(\Gamma)\,=\, \prod_f\int_{\mathbb{R}^3}dx_f\prod_L\int_{\SU(2)}dh_L\,\prod_f\, e^{i\frac{1}{2}Tr(x_f G_f)}\,*\,\delta_{G_f^{-1} x_f G_f} (x_f),
\ee
with $G_f = \prod_{L\in\partial f} h_L$, i.e. we obtain exactly the BF (3d gravity) simplicial path integral, by identification of the Lie algebra variables $x_f$  with the discrete triad associated to each edge: $x_f \equiv B_f$, with an additional constraint $\delta_{G_f^{-1} B_f G_f} (B_f)$, whose commutative analogue would impose that the holonomy $G_f$ lies in the plane orthogonal to the edge to which $B_f$ is associated (the implications of the non-commutative constraint, on the other hand, are still being investigated). 

\

\noindent The same construction goes through with the same result also in the 4d case, for BF theory. It is then clear that this new non-commutative representation of GFT is a representation in terms of metric variables (as opposed to connection variables, and that it gives Feynman amplitudes which have directly the form of simplicial path integrals.

\

\noindent To construct a corresponding gravity model in 4d, one has then to impose the gravity constraints on the 4d GFT, once this is written as a non-commutative field theory on the Lie algebra $\so(4)^4$. They can be imposed on the Lie algebra variables $B_f$ labeling the 2-cells $f$ of $\Gamma$, dual to triangles of the simplicial complex $\Delta$, requiring that  $\exists k_t\in S^3\sim\SU(2)$ (normal to tetrahedron $t$), s.t. $B_f^- = - k_t^{-1} B_f^+ k_t \;\;\;\;\;\forall f\subset t$, and that $\sum_{f\subset t} B_f = 0$. In turn, these two conditions can be imposed by means of the two projection operators $ S_k \;\equiv \;\prod_{f\subset t} \delta_{-k_t^{-1} B_f^+ k_t}(B_f^-) $ and
$ C \;\equiv\; \delta\left( \sum_{f\subset t} B_f \right) $ acting on each field $\phi$ in the GFT action.
In order to implement the correct simplicial geometry, it turns out to be necessary to generalize slightly the standard definition of the field to $\phi(B_1,B_2,B_3,B_4; k )$, with $k\in S^3$, with extended invariance: $\phi(B_1,B_2,B_3,B_4; k ) = \phi( h  B_1, h B_2, B_3, h B_4; h^{-1} k )$ with $h\in \Spin(4)$. On this field the two projections $S_k$ and $C$ can be imposed giving: 
$\Phi(B_1^+,B_2^+,B_3^+,B_4^+;k)\,=\left(S_{k}\,*\,C\,*\,\phi\right)(B_1,B_2,B_3,B_4; k)$. 

\

\noindent Various models can be defined, and the identification of the correct way of imposing them and the definition of the \lq correct\rq GFT model (with and without the Immirzi parameter) is in progress and will be reported elsewhere \cite{ioaristide2}. In any case all the models being considered within this formalism and imposing the above constraints produce Feynman amplitudes with the general form:
\be
Z(\Gamma)\,=\,\prod_f\int_{\so(4)}dB_f\prod_t\int_{\Spin(4)}dh_L\,\prod_f \left(  S_1\,*\,S_2\,\right)(B_f)\,*\,e^{\frac{i}{2}Tr (B_f G_f)}. \ee
Thus they are again nice simplicial gravity path integrals, with manifest implementation of the simplicial geometry, for a BF theory on which one has imposed both the gravity constraints $S_1$ on the Lie algebra variables $B_f$,  and  secondary constraints $S_2 $ on the connection variables $h_L$, ensuring consistency between the gravity constraints and the parallel transport. This 4d construction and the analysis of the corresponding model(s) is currently being completed \cite{ioaristide2}.

\

\noindent It is anyway clear that this new GFT representation realizes an explicit duality between simplicial quantum gravity and loop quantum gravity (spin foam models), on top of bringing GFTs in close contact with non-commutative geometry.

\subsection{Group field theory renormalization and the sum over topologies}
\noindent A second area of recent developments \cite{ren,borel} has been the application of quantum field theory techniques to GFTs, to gain a better understanding and control over its perturbative expansion, using tools from renormalization theory. 

\

\noindent GFTs define a sum over simplicial complexes 1) of arbitrary topology and 2) that correspond, in general, to pseudo-manifolds, i.e. contain conical singularities at the vertices. The issue of controlling the sum over topologies, and of identifying an approximation in which simple topologies dominate, has an analogue in the context of matrix models \cite{mm}. In matrix models, it has been shown that, in the so-called large-N limit, diagrams of trivial topology ($S^2$ in the compact case) dominate the perturbative sum. The issue of controlling the relative weight of manifolds and pseudo-manifolds in the perturbative sum arises instead only in dimensions $D>2$ and it has represented an obstacle to the development of tensor models. A third issue is to identify and control the divergences that arise in this pertrbative expansion, which are of two types: a) divergences in the sum over (pre-)geometric data (group elements or group representations) for each amplitude associated to a given simplicial complex; b) the divergence of the entire sum over simplicial complexes. 

\

\noindent The work of \cite{ren} makes the first steps toward solving these three issues, leaving aside 3b), starting a systematic study of GFT renormalization, in the context of the Boulatov model for 3d (Riemannian) quantum gravity, that we have discussed above. 
The divergences of this model are related to the topology of the bubbles (3-dimensional cells), dual to vertices of the simplicial complex, in the Feynman diagrams, but it is difficult to establish which diagrams need renormalization in full generality, mainly due to the very complicated topological structure of $3D$ simplicial complexes, after a scale is introduced in the theory by an explicit cut-off in the spectral decomposition of the propagator.

\noindent What is achieved in \cite{ren} is the following: 

\begin{itemize}
\item  a detailed algorithm is given for identifying bubbles (3-cells) in the Feynman diagrams of the model, together with their boundary triangulations, which in turn can be used to identify the topology (genus) of the same boundary.

\item using this algorithm, the authors are able to identify a subclass of Feynman diagrams which allow for a complete contraction procedure, and thus the ones that allow for an almost standard renormalization; moreover, this class of graphs, dubbed \lq\lq type 1\rq\rq is shown to be a natural generalization of the 2d planar graphs of matrix models, thus suggesting that they can play a similar role in GFTs to that of planar diagrams in matrix models.
Notice also that the existence of such contraction procedure can be seen as a sort of generalised locality property

\item for this class of diagrams, an exact power counting of divergences is proven, according to which their divergence is of the order:

$$ A_{\Gamma}\,=\,\left(\delta^{\Lambda}(I)\right)^{|{\cal B}_{\Gamma}|-1}$$
where $| {\cal B}_\Gamma|$ is the number of bubbles in the diagram $\Gamma$, and $\delta^\Lambda(I)$ is the delta function on the group, with representation cut-off  $\Lambda$, evaluated at the identity $I$.

\end{itemize}

\noindent On the basis of these results, of the experience gained with esplicit evaluation of Feynman amplitudes, and of a better understanding of the combinatorial structure of the Feynman diagrams, it was then possible to put forward two main conjectures, obviously confirmed in all examples considered: 1) that {\it all} \lq\lq type 1\rq\rq diagrams correspond to {\it manifolds} of {\it trivial topology}; and 2) that an appropriate generalization of the usual scaling limit (large-N) of matrix models to these GFT would lead to the relative suppression of all the \lq\lq non-type 1\rq\rq diagrams, and thus leave us with: only type 1 diagrams in need for renormalization, and only manifolds of trivial topology in the theory.

\

\noindent A different perspective on GFT divergences is taken in \cite{borel}, which also tackles the difficult issue of the summability of the entire perturbative sum (thus including the sum over topologies). The authors consider both the Boulatov model and a modification of the same proposed in \cite{freidellouapre}, obtained adding a second interaction term in the action:

\begin{equation}
+\frac{\lambda\,\delta}{4!}\prod_{i=1}^{6}\int dg_i\left[
\, \phi(g_1,g_2,g_3)\phi(g_3,g_4,g_5)\phi(g_4,g_2,g_6)\phi(g_6,g_5,g_1)\right].
\end{equation}

\noindent The new term corresponds  to the only other possible way of gluing 4 triangles to form a closed surface. This mild modification gives a Borel summable partition function \cite{freidellouapre}. This shows that a control over the sum over topologies and a non-perturbative definition of the corresponding GFT is feasible.

\noindent For both the Boulatov model and the modified one, the authors of \cite{borel} establish general perturbative bounds on amplitudes using powerful constructive techniques, rather than focusing on explicit power counting or Feynman evaluations. They find that, using the same regularization used in \cite{ren}, the amplitudes of the Boulatov model for a diagram with $n$ vertices, are bounded, with cut-off $\Lambda$, by $K^n \Lambda^{6+3 n/2}$, for some arbitrary positive constant $K$, while the modifed model has amplitudes bounded by $K^n \Lambda^{6+3 n}$, and that both bounds can be saturated. This result shows that the Freidel-Louapre modification (BFL), even though Borel summable, is perturbatively more divergent that the original model.

\noindent The second main result of \cite{borel} relies again on constructive field theory techniques. A cactus expansion of the BFL model is obtained, and used to prove  the Borel summability of the free energy of the model and to define its Borel sum. We can expect more applications of these techniques to other GFT models, also in higher dimensions.  

\

\noindent Similar techniques have been also applied in \cite{JosephVincent} to obtained perturbative bounds, in any dimensions, on the colored GFT models for BF theory first defined in \cite{razvan}. This colored extension of the GFT formalism is particularly suited for the analysis of the topology of the GFT Feynman diagrams, in that, for example, the bubbles are readily identified, and has been used in \cite{razvan} to 1) show that the associated Feynman diagrams are cellular complexes; 2) define their cellular homology; 3)  define a related homotopy transformation; 4) relate the GFT amplitudes for these diagrams to the fundamental group;  5) prove that the corresponding closed \lq\lq Type 1 graphs\rq\rq are homotopically trivial, and thus, by the Poincar\'e conjecture, 3-spheres. This proves the first conjecture put forward in \cite{ren}.
Moreover, in \cite{razvan2} the boundary graphs of the same colored models are identified, and the topological (Bollobas-Riordan) Tutte polynomials associated to (ribbon) graphs are generalized to topological polynomials adapted to colored group field theory diagrams in arbitrary dimension.

\subsection{Emergent non-commutative matter fields from group field theories}
\noindent The last set of results we want to mention are interesting steps in the direction of bridging the gap between the microscopic GFT description of quantum space (and the language of spin networks, simplices, spin foams, etc) and macroscopic continuum physics, including usual General Relativity and quantum field
theories for matter. In fact, this problem is faced by {\it all} discrete approaches
to quantum gravity\cite{libro}.

\

\noindent One would expect \cite{fluid} a generic continuum spacetime to be formed by
zillions of Planck size building blocks, and thus to be, from the GFT point of view, a
many-particle system whose microscopic theory is given by some
fundamental GFT action. This suggests us to look for
ideas and techniques from statistical field theory and condensed
matter theory, and to try to apply/re-interpret them
in a GFT context. 

\noindent Condensed matter theory also provides examples of systems in which the
collective behaviour of the microscopic constituents, in some
hydrodynamic approximation, gives rise to effective emergent
geometries from the collective variables themselves\cite{analog}. What happens is that
the collective parameters describing the fluid and its dynamics can be recast as the component
functions of an {\it effective metric field}, and that the effective dynamics of perturbations (quasi-particles, themselves
collective excitations of the fundamental constituents of the
fluid) takes the form of matter field theories in curved spacetimes, on the same effective metrics.

\

\noindent Inspired by these results, we ask: assuming that a given GFT model describes
the microscopic dynamics of (the fundamental building blocks of) a
{\it discrete quantum spacetime}, and that some solution of the
corresponding fundamental equations can be interpreted as
identifying a given quantum spacetime configuration, 1) can we
obtain an effective macroscopic {\it continuum} field theory for
matter fields from it? and if so, 2) what is the effective
spacetime and geometry that these emergent matter fields see?

\

\noindent As it turns out, it is possible \cite{emergentmatter} to apply the same procedure to GFT
models and obtain rather straightforwardly effective
continuum field theories for matter fields. The effective matter
field theories that we obtain most easily are QFTs on non-commutative spaces of Lie algebra type.

\

\noindent The basic point is the use of the same natural duality between Lie algebra and corresponding Lie group, that we have exploited  in the previous section, to construct a new (non-commutative) representation for GFTs, now re-interpreted as the non-commutative version of the usual duality between coordinate and momentum space. More precisely, if we have a non-commutative spacetime of Lie algebra type $[X_\mu, X_\nu] = C_{\mu\nu}^\lambda X_\lambda$, the corresponding momentum space is naturally identified with the corresponding Lie group, in such a way that the non-commutative coordinates $X_\mu$ act on it as (Lie) derivatives (as we expect in the quantum setting). The link with GFTs is then obvious: in momentum space the field theory on such non-commutative spacetime will be given, by definition, by some sort of group field theory. The task will then be to derive the relevant field theories from matter from interesting GFT models of quantum spacetime. 

\

\noindent In 3 spacetime dimensions the results obtained recently \cite{phase} concern the euclidean signature and a non-commutative spacetime given by the $\su(2)$ Lie algebra . 

\noindent The group field theory we start from is again the Boulatov model.
We look \cite{phase} at two-dimensional variations $\delta\phi(g_1,g_2,g_3)\equiv\psi(g_1g_3^{-1})$ around a class of GFT classical solutions (they can be interpreted as quantum flat space on some a priori non-trivial topology) given by: \be
\phi^{(0)}(g_1,g_2,g_3)\,=\,\sqrt{\frac{3!}{\lambda}}\,\int dg\;
\delta(g_1g)F(g_2g)\delta(g_3g), \quad F:G\rightarrow\R, \quad \int F^2=1 \ee

\noindent The effective action is then:
\bes
S_{eff}[\psi]&\equiv&\,S_{3d}[\phi^{(0)}+\psi]-S_{3d}[\phi^{(0)}]\,=\,  \frac{1}{2}\int\psi(g)\kk(g)\psi(g^{-1})- \nonumber \\ &-& \frac{\mu}{3!}\int[dg]^3\,
\psi(g_1)\psi(g_2)\psi(g_3)\delta(g_1g_2g_3)
-\frac{\lambda}{4!}\int[dg]^4\, \psi(g_1)..\psi(g_4)\delta(g_1..g_4),\nonumber
\ees with the kinetic term and the 3-valent coupling given in term
of $F$:
$$
\kk(g)\,=\,1-2\left(\int F\right)^2-\int dh F(h)F(hg), \qquad
\frac{\mu}{3!}\,=\,\sqrt{\frac{\lambda}{3!}}\,\int F.
$$
with $F(g)$ assumed to be invariant under conjugation $F(g)=F(hgh^{-1})$. 
This is a non-commutative quantum field theory invariant
under the quantum double of $\SU(2)$ (which provides a quantum
deformation of the Poincar\'e group $\ISU(2)$).

\noindent Expanding $F$ in group characters: $ F(g)=\sum_{j\in\N/2}
F_j\chi_j(g)$, where $j\in\N$ label irreducible representations of $\SU(2)$,
the kinetic term reads: \be \kk(g)=1-3F_0^2-\sum_{j\ge 0}\frac{F_j^2
}{d_j} \chi_j(g)=\sum_{j\ge 0}
F_j^2\left(1-\frac{\chi_j(g)}{d_j}\right)-2F_0^2\,\equiv\,
Q^2(g)-M^2. \ee  We can interpret $Q^2(g)\ge 0$ as a generalized ``Laplacian", and $F_0^2$ as a \lq\lq gravitational\rq\rq mass $M^2$.
For the simple classical solution $F(g)=a+\sqrt{1 - a^2} \chi_1(g)$, we obtain \be
\kk(g)=\frac{4}{3}(1-a^2)\,\vec{p}^2 -2a^2. \ee 

\

\noindent Similar results have also been obtained in the 4d context \cite{gftdsr,emergentmatter}. It has been shown that from GFT models (indirectly) related to 4d quantum gravity, it is possible to derive effective non-commutative matter field theories of \lq\lq deformed special relativity\rq\rq type, based on momentum group given by $\AN(3)\approx SO(4,1)/SO(3,1)$ and a non-commutative $\ka$-Minkowski spacetime: $[x_0,x_i] = i \ka x_i$; these field theories form the basis for much current work in the area of quantum gravity phenomenology \cite{QGPhen}.

\

\noindent Work in this direction, therefore, including these recent results, is a step in bridging the gap between the
fundamental discrete theory of spacetime we have at hand, and a
continuum description of spacetime, and getting closer to possible quantum gravity
phenomenology, thus bringing this class of models a bit closer to
experimental falsifiability. Let us also notice that, contrary to the situation in analog gravity mdoels in condensed matter, we have
here models which are non-geometric and far from usual
geometrodynamics in their formalism, but which at the same time
are expected to encode quantum geometric information and indeed to
determine, in particular in their classical solutions, a (quantum
and therefore classical) geometry for spacetime \cite{iogft}. We
are, in other words, far beyond a pure analogy.

\section{Conclusions}
\noindent We have introduced the key ideas behind the group field theory approach to quantum gravity, and to the microscopics of quantum space, and the basic elements of its formalism. We have also briefly reported on some recent results obtained in this approach, concerning both the mathematical definition of these models, and  possible avenues towards extracting interesting physics from them. From our outline it should be clear that, while much more work is certainly needed in this area of research, the new direction toward quantum gravity that group field theories provide is exciting and full of potential.

\begin{theacknowledgments}
 We thank the organizers of the Max Born symposium, in particular Jurek Kowalski-Glikman, and all the attendees, for a very stimulating and enjoyable meeting.
\end{theacknowledgments}

\end{document}